# Origin of the metamagnetic transitions in $Y_{1-x}Er_xFe_2(H,D)_{4.2}$ compounds

V. Paul-Boncour[1], O. Isnard[2,3], V. Shtender[1], Y. Skourski[4], M. Guillot[2,5]

[1]*Université Paris-Est Créteil, CNRS, ICMPE-UMR7182, F-94320 Thiais, France*
[2]*Univ Grenoble Alpes, 25 Rue Martyrs, F-38042 Grenoble, France*
[3]*CNRS, Inst NEEL, 25 Rue Martyrs, F-38042 Grenoble, France*
[4]*Helmholtz Zentrum Dresden Rossendorf, Hochfeld Magnetlabor Dresden (HLD-EMFL), D-01328 Dresden, Germany*
[5]*CNRS, Laboratoire National de Champs Magnétiques Intenses (LNCMI-EMFL), 25 rue des Martyrs, BP166, F-38042 Grenoble 9, France*

**Abstract**

The structural and magnetic properties of $Y_{1-x}Er_xFe_2$ intermetallic compounds and their hydrides and deuterides $Y_{1-x}Er_xFe_2H(D)_{4.2}$ have been investigated using X-ray diffraction and magnetic measurements under static and pulsed magnetic field up to 60 T. The intermetallics crystallize in the C15 cubic structure (*Fd-3m* space group), whereas corresponding hydrides and deuterides crystallize in a monoclinic structure (*Pc* space group). All compounds display a linear decrease of the unit cell volume *versus* Er concentration; the hydrides have a 0.8% larger cell volume compared to the deuterides with same Er content. They are ferrimagnetic at low field and temperature with a compensation point at $x = 0.33$ for the intermetallics and $x = 0.57$ for the hydrides and deuterides. A sharp first order ferromagnetic-antiferromagnetic (FM-AFM) transition is observed upon heating at $T_{FM-AFM}$ for both hydrides and deuterides. These compounds show two different types of field induced transitions, which have different physical origin. At low temperature ($T < 50$ K), a forced ferri-ferromagnetic metamagnetic transition with $B_{trans1} \approx 8$ T, related to the change of the Er moments orientation from antiparallel to parallel Fe moment, is observed. $B_{trans1}$ is not sensitive to Er concentration, temperature and isotope effect. A second metamagnetic transition resulting from antiferromagnetic to ferrimagnetic state is also observed. The transition field $B_{trans2}$ increases linearly versus temperature and relates to the itinerant electron metamagnetic behavior of the Fe sublattice. An onset temperature $T_{M0}$ is obtained by extrapolating $T_{FM-AFM}(B)$ at zero field. $T_{M0}$ decreases linearly versus the Er content and is 45±5 K higher for the hydrides compared to the corresponding deuteride. The evolution of $T_{M0}$ versus cell volume shows that it cannot be attributed exclusively to a pure volume effect and that electronic effects should also be considered.

# I. Introduction

The magnetic properties of intermetallic compounds with rare earth ($R$) and 3d transition metal ($T$) have been extensively studied as they as they offer the opportunity to mix the localized magnetism of $R$ metal with the itinerant magnetism of 3d transition metal [1]. These compounds can present fascinating magnetic properties due to the unique combination of large magnetic moments and magnetic anisotropy arising from $R$ and elevated transition temperatures originating from $T$. Such high performance magnetic properties are very interesting for many applications like permanent magnets, magnetocaloric compounds [2-7].

Insertion of atoms of light elements as H, C, or N atoms has been used to tune the magnetic properties of such $RT_n$ intermetallic compounds [8-12]. Many studies have been undertaken on the magnetic properties of hydrides showing, for example, that the magnetic ordering temperature can be increased by H insertion in $R$Mn$_2$ [13-18] or $R_2$Fe$_{17}$ [19-21] compounds and decreased in $R$Fe$_2$ [22-28] or $R_6$Mn$_{23}$ [29, 30] compounds. The possibility to tune the magnetic properties by H insertion has raised interest not only for fundamental properties but also for applications. For instance, it can improve the properties of permanent magnets, shift the Curie temperature of La(Fe$_{1-x}$Si$_x$)$_{13}$ compounds near room temperature for magnetic refrigeration [31, 32].

Hydrogen insertion in $RT_n$ compounds leads to a large cell volume increase, a modification of the electronic properties and in some cases a lowering of the crystal symmetry which all influence the magnetic interactions. The magnetism of $T$ elements, related to direct 3d-3d interactions, is strongly influenced by the modification of the density of state (DOS) at the Fermi energy ($E_F$), which determines the onset of ferromagnetism. The interactions between the 4f magnetic moments of the $R$ elements, occurs through an indirect exchange via the 5d conduction electrons and are generally lowered by the cell volume increase due to the decrease of the overlap between the electronic orbital of neighboring atoms. The interaction between $R$ and $T$ moments occurs also via an indirect exchange mechanism between the 3d and 5d orbitals and is also reduced when the distances between $R$ and $T$ atoms increase as the 5d-3d overlap and therefore their hybridization becomes weaker [33-36]. As a result of the above indirect interaction mechanisms the $R$ and $T$ moments are coupled parallel for light $R$ and antiparallel for heavy $R$ as for the intermetallic compounds.

The study of $R$Fe$_2$ hydrides is of particular interest to study the change of all these magnetic interactions as these compounds can absorb a large amount of hydrogen, up to 5 H/f.u., and form intermediate hydrides with different H contents and crystal structures [23, 25, 27, 37-45]. In all these compounds the magnetic ordering temperature decreases *versus* H content. Rhombohedral distortions observed for H content larger than 3.5 were also found to influence the magnetocrystalline anisotropy of the $R$ moments [25].

The investigations of ErFe$_2$ hydrides have shown a strong decrease of the compensation point from 410 to 42 K as the H concentration increases from 0 to 3.9 H/f.u. due to the reduction of the Er and Fe interactions [24, 38, 46]. A rise of the Fe moment is observed for a hydrogen uptake up to 3.3 H/f.u. [45, 47-49], then it is reduced for larger H concentration. The Er moment measured by neutron diffraction is also reduced upon H absorption compared to the free ion value (9 $\mu_B$/Er) observed in ErFe$_2$ [38, 46, 50, 51]. A loss of ordered Fe moment and a canting of the Er moments below 5 K was found in ErFe$_2$D$_5$ [51].

The study of the magnetic properties of YFe$_2$ hydrides showed, as in ErFe$_2$ hydrides, that at 4.2 K the mean Fe moment increases up to 3.5 H/f.u. and decreases for larger H content up to 5 H/f.u. where the Fe moments are no more ordered, YFe$_2$H$_5$ being paramagnetic [44]. First principles calculations revealed that this results from a competition between the hydrogen induced magnetovolume effect which increases the Fe moment by a narrowing of the 3d band and the strong Fe-H bonding which reduces the Fe-Fe interaction and consequently the Fe moment for H content larger than 3.5 H/f.u. [52-54].

In YFe$_2$H$_{4.2}$, H insertion into preferential tetrahedral Y$_2$Fe$_2$ and YFe$_3$ interstitial sites induces a monoclinic distortion (*Pc* space group) [55]. This distortion generates a strong structural anisotropy and different number of H neighbors around each of the 8 Fe sites in the cell. A sharp first order ferromagnetic-antiferromagnetic transition (FM-AFM) observed at $T_{FM-AFM}$ = 131 K for YFe$_2$H$_{4.2}$ [56]. This transition displays the characteristics of an itinerant electron metamagnetic (IEM) behavior of the Fe sublattice as observed in hexagonal $A$Fe$_2$ compounds ($A$ = Hf, Ta) [57-61]. A metamagnetic behavior is observed above $T_{FM-AFM}$, with transition field increasing linearly versus temperature. $T_{FM-AFM}$ is very sensitive to any structural change, and a very unusual giant (H, D) isotope effect has been observed for this system. A shift of 47 K (50%) of $T_{FM-AFM}$ has been found between YFe$_2$D$_{4.2}$ and YFe$_2$H$_{4.2}$ [56]. As the hydride has a larger cell volume than the deuteride ($\Delta V/V$ = 0.8%), the $T_{FM-AFM}$ shift was attributed to a strong magnetovolume effect on the IEM behavior of the Fe sublattice. This can be understand in relation with the electronic properties of YFe$_2$H$_{4.2}$: as $E_F$ is located on the slope of a sharp DOS peak, a very small cell volume variation changing the width of this peak strongly modifies the DOS intensity at $E_F$ and therefore the stability of the ferromagnetic state. The sensitivity of $T_{FM-AFM}$ to the cell volume variation was confirmed by the influence of an external pressure on the magnetic properties of both hydride and deuteride: $T_{FM-AFM}$ decreases linearly and is suppressed for pressure larger than 0.5 GPa and 1.2 GPa for YFe$_2$D$_{4.2}$ and YFe$_2$H$_{4.2}$ respectively [62, 63]. Above these pressure and at low temperature the ground state becomes antiferromagnetic [62]. In addition, the cell volume variation versus pressure was measured for both hydride and deuteride and coupled to the variation of $T_{FM-AFM}$ versus pressure, showing the existence of the same critical cell volume for the onset of ferromagnetism [62, 63].

Another way to tune the cell volume of YFe$_2$H(D)$_{4.2}$ compounds is to substitute Y by another $R$ element like Gd, Tb, Er or Lu as reported in [64]. The structural and magnetic properties of pseudobinary Y$_{1-x}$Er$_x$Fe$_2$D$_{4.2}$ compounds ($x$ = 0.3 and 0.5) have been already studied by combining high magnetic field and neutron diffraction experiments [65-67]. Both compounds display the same nuclear structure than YFe$_2$D$_{4.2}$ with a disordered Er for Y substitution and a cell volume reduction attributed to the smaller size of the Er atom. The decrease of $T_{FM-AFM}$ as a function of Er concentration, was therefore related to this cell volume contraction. These studies revealed two types of metamagnetic transitions: a forced ferri-ferromagnetic transition (Ferri-FM) at low temperature and an AFM-FM transition above $T_{FM-AFM}$ which depends of the Er concentration and the nature of the hydrogen isotope. The first transition has been related to a lowering of the Er-Fe exchange interactions by H(D) absorption, whereas the second one was, like for the YFe$_2$(H,D)$_{4.2}$ compounds, attributed to the IEM behavior of the Fe sublattice.

The aim of this new study is to investigate systematically the influence of Y for Er substitution combined to the (H, D) isotope effect on these different metamagnetic transitions using complementary magnetic experiments with steady and pulsed magnetic fields allowing to reach a maximum field of 60 T. The magnetic properties of the parent intermetallic compounds have been also presented as references. The first purpose of this work is to understand why the transition field of the ferri-ferromagnetic transition is so small compared to most other rare-earth and transition metal compounds which present forced ferri-ferromagnetic transitions above 20 T [68-71]. A second purpose was to verify if the (H, D) isotope substitution, the external and chemical pressure influence on the FM-AFM transition are only due to a magnetovolume effect or if other parameters should be considered. This will allow to understand more deeply the physical origin of this second metamagnetic transition.

## II. Experimental

The synthesis of the Y$_{1-x}$Er$_x$Fe$_2$ intermetallic compounds ($x$ = 0, 0.1, 0.3, 0.5, 0.7 and 1) and their corresponding hydrides and deuterides is described in details in refs [64, 72]. The intermetallics were prepared by induction melting and annealed 4 weeks at 1073 K. The hydrides and deuterides were prepared by solid-gas reaction using a Sieverts apparatus. To avoid H or D desorption their surface was passivated by a process described in [73]. The samples were kept in liquid nitrogen tank to avoid large desorption versus time.

The crystal structure of all the compounds was measured by powder X-ray diffraction at room temperature on a D8 Bruker diffractometer (Cu K$_\alpha$ radiation). The XRD patterns were refined using the Fullprof Suite code [74].

Magnetization measurements were performed on all intermetallic compounds as well as hydrides and deuterides with 4.2 ± 0.2 H(D)/f.u. using a conventional Physical Properties Measurement System (PPMS-9T) from Quantum Design. Magnetization measurements were performed for selected hydrides and deuterides in the Grenoble High Magnetic Field Laboratory (LNCMI) located at Grenoble in France in high static magnetic field up to 35 T. The experimental procedures of the magnetic measurements are detailed in [67]. Isothermal variations $M_T(B)$ were then obtained. It is noted that after sweeping the field up and down at a given temperature the absence of remanent magnetization was checked before heating. The iso-field behavior was then deduced from the isothermal variations. Additional magnetic measurements on selected deuterides were performed under pulsed magnetic fields up to 60 T at 4.2 K in the Dresden High Magnetic Field Laboratory (HLD) in Germany. The magnetization curves were scaled using the previous experimental data.

## III. Experimental results

### A. Intermetallic compounds

The structural and magnetic properties of the $Y_{1-x}Er_xFe_2$ intermetallic compounds have been measured for $x = 0, 0.1, 0.3, 0.5, 0.7$ and 1. The XRD pattern of $Y_{0.9}Er_{0.1}Fe_2$ is presented as an example in Figure 1. All these compounds crystallize in the C15 cubic structure and their cell volume decreases (Table 1) linearly versus the Er content according to a Vegards law with a slope $\Delta V/\Delta x = -12.64(3)$ Å$^3$ as a consequence of the smaller size of Er atom as compared to Y one.

**Table 1**: Cell parameters measured at room temperature of $Y_{1-x}Er_xFe_2$ compounds

| Compound | $a$ (Å) | $V$(Å$^3$) |
|---|---|---|
| YFe$_2$ | 7.357(1) | 398.20(3) |
| Y$_{0.9}$Er$_{0.1}$Fe$_2$ | 7.349(1) | 396.90(3) |
| Y$_{0.7}$Er$_{0.3}$Fe$_2$ | 7.334(1) | 394.48(3) |
| Y$_{0.5}$Er$_{0.5}$Fe$_2$ | 7.318(1) | 391.90(3) |
| Y$_{0.3}$Er$_{0.7}$Fe$_2$ | 7.303(1) | 389.50(3) |
| ErFe$_2$ | 7.285(1) | 386.62(3) |

ErFe$_2$ is a ferrimagnet with a Curie temperature $T_C$ of 590 K and a compensation point at 490 K [50]. Neutron powder diffraction NPD study indicated that both Er and Fe sublattices order magnetically at 590 K. The dilution of Er by Y substitution induces a linear decrease of $T_C$ from 590 to 545 K [75].

The magnetization curves versus temperature $M(T)$ (Figure 2a) are characteristic of a ferrimagnetic behavior. A minimum in magnetization is observed for $x = 0.1$ to 0.5, much below 300 K, bearing witness of the existence of compensation temperatures ($T_{comp}$), that is temperature at which the Er and Fe sublattice magnetization cancel out. These minima are above room temperature for $x = 0.7$ and 1. The compensation temperatures are relatively more reduced than the Curie temperature as the Y content increases, this has been explained by a lowering of the Er moment ordering temperature compared to that of the Fe [75].

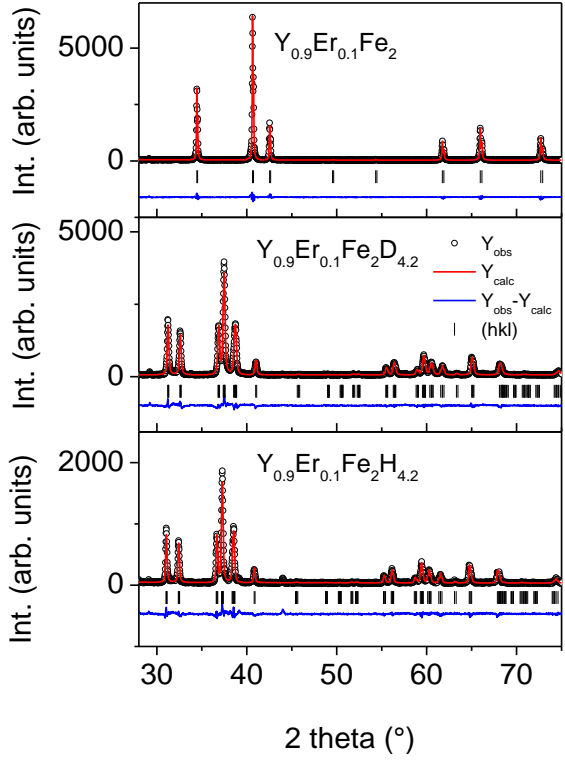

**Figure 1**: XRD patterns of $Y_{0.9}Er_{0.1}Fe_2$ alloy refined in a cubic $Fd\text{-}3m$ space group, and corresponding hydride and deuteride refined in a monoclinic $C2/m$ space group ($\lambda$ = Cu $K_\alpha$). The weak additional peaks are due to $Y_2O_3$.

The experimental saturation magnetization $M_{Sat}$ at 5 K, decreases for $x \leq 0.3$ and increases again for larger Er content (Figure 1b). Assuming that the Fe moments are collinear and antiparallel to the Er moments as observed by neutron powder diffraction (NPD) for $ErFe_2$, and taking into account a small magnetic moment induced on Y site [52], an average moment $M_{calc}$ was obtained by fitting the experimental values according to:

$$M_{sat} = | 2M_{Fe} - xM_{Er} - (1-x) M_Y | \tag{1}$$

with $M_{Fe}$ = 1.6 $\mu_B$, $M_{Er}$ = 9 $\mu_B$, $M_Y$ = 0.34 $\mu_B$, $\chi^2$ = 4.5 %

$M_{sat}$ minimum is located at $x$ = 0.33 and corresponds to the compensation point when the (Er, Y) and Fe magnetic sublattice magnetizations cancel out. The magnetization curves versus field $M_{5K}(B)$ present hysteresis loops with a significant variation of the remanent magnetization ($M_R$) and the coercive field ($B_C$) versus Er concentration (Figure 2b). $B_C$ is maximum at $x$ = 0.3, i.e. when the two magnetic sublattices compensate each other. These results are in good agreement with those of Alves et al. (1994) [76] and are explained by intrinsic domain wall pinning. A larger applied field is required to reverse the magnetization near the compensation point. The remanent magnetization $M_R$ is also sensitive to the Er content, with minima at $x = 0$, 0.3 and 1 (Figure 2b), i.e. for the binaries and close to the compensation point.

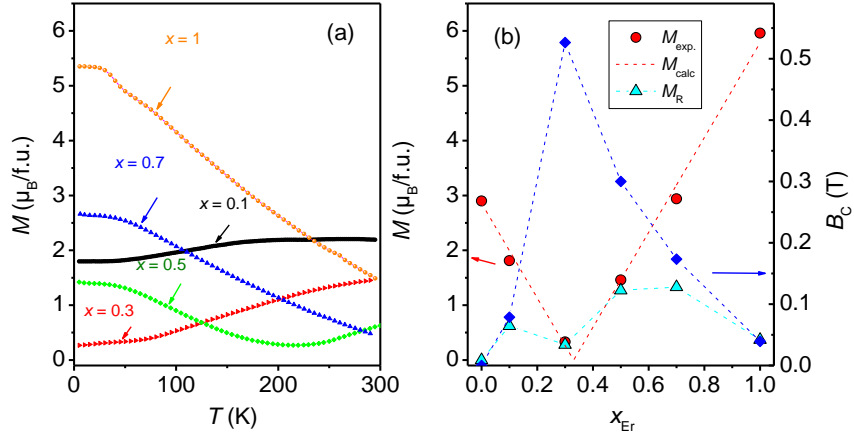

**Figure 2**: (a) Thermomagnetization curves ($B = 1.5$ T) (b) experimental ($T = 5$ K) and calculated saturation magnetization $M_{exp}$ and $M_{calc}$, remanent magnetization $M_R$ and coercive field $B_C$ of $Y_{1-x}Er_xFe_2$ intermetallic compounds.

## B. Hydrides and deuterides

### 1. Structural properties

The thermodynamic properties of $Y_{1-x}Er_xFe_2$ hydrides and deuterides have been already described in ref [64]. For an H (D) content of about 4.2 H(D)/f.u., these compounds crystallize in the same monoclinic structure as $YFe_2H_{4.2}$ and $YFe_2D_{4.2}$ [55]. This structure results from a distortion of the $C15$ cubic structure induced by hydrogen or deuterium order into specific tetrahedral interstitial sites ($R_2Fe_2$ and $RFe_3$) [55, 77]. The XRD patterns can be refined in a monoclinic structure with a $C2/m$ space group as presented in Figure 1 for the hydride and deuteride of $Y_{0.9}Er_{0.1}Fe_2$. However, the previous neutron diffraction study of the deuterides with $x = 0$, 0.3 and 0.5 revealed a lowering of the symmetry in a $Pc$ space group with a doubling along the ***b*** axis due to D order [55, 73, 78]. For comparison with these previous works, we have chosen to keep the crystal structure description in the $Pc$ space group. The synthesis of $ErFe_2H_{4.2}$ was not successful and only $ErFe_2D_{4.2}$ properties are presented in this work. The general tendency is a decrease of the *a, b,* and *c* cell parameters and *V* cell volume versus Er content and systematic larger cell volumes for the hydrides compared to the corresponding deuterides (Table 2). The monoclinic angle beta is between 122.35 and 122.65° and does not change significantly versus Er concentration. The cell volumes are fitted with a linear equation $V_H = 506.8-13.x$ and $V_D = 503.6-13.x$ corresponding to a cell volume variation by Er atom of -2.53% and -2.58% for the hydride and deuteride respectively (Figure 3).

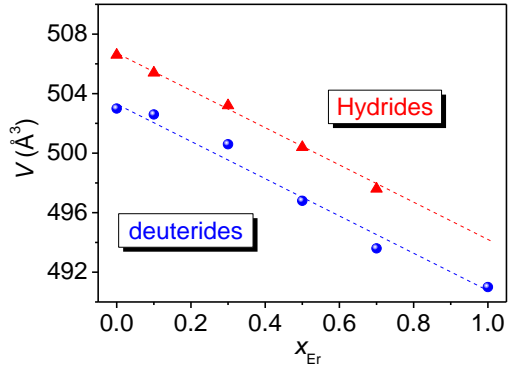

**Figure 3**: Cell volume variation of $Y_{1-x}Er_xFe_2$ hydrides and deuterides versus Er content.

**Table 2:** Cell parameters measured at room temperature of the indicated $RFe_2$ hydrides and deuterides

|  | $a$ (Å) | $b$ (Å) | $c$ (Å) | $\beta$ (°) | $V$ (Å³) |
|---|---|---|---|---|---|
| $YFe_2H_{4.2}$ | 5.515(1) | 11.474(2) | 9.445(1) | 122.401(1) | 506.8(1) |
| $Y_{0.9}Er_{0.1}Fe_2H_{4.2}$ | 5.515(1) | 11.486(2) | 9.440(1) | 122.361(1) | 505.30(4) |
| $Y_{0.7}Er_{0.3}Fe_2H_{4.2}$ | 5.513(1) | 11.482(2) | 9.422(1) | 122.459(1) | 503.18(2) |
| $Y_{0.5}Er_{0.5}Fe_2H_{4.2}$ | 5.496(1) | 11.456(2) | 9.409(1) | 122.345(1) | 500.48(4) |
| $Y_{0.3}Er_{0.7}Fe_2H_{4.2}$ | 5.494(1) | 11.428(2) | 9.395(1) | 122.488(1) | 497.56(9) |
| $YFe_2D_{4.2}$ | 5.507(1) | 11.472(2) | 9.428(1) | 122.354(1) | 502.96(9) |
| $Y_{0.9}Er_{0.1}Fe_2D_{4.2}$ | 5.506(1) | 11.468(2) | 9.429(1) | 122.397(1) | 502.68(2) |
| $Y_{0.7}Er_{0.3}Fe_2D_{4.2}$ | 5.492(1) | 11.450(2) | 9.411(1) | 122.355(1) | 499.96(4) |
| $Y_{0.5}Er_{0.5}Fe_2D_{4.2}$ | 5.482(1) | 11.428(2) | 9.390(1) | 122.374(1) | 496.80(9) |
| $Y_{0.3}Er_{0.7}Fe_2D_{4.2}$ | 5.481(1) | 11.408(2) | 9.377(1) | 122.644(1) | 493.66(4) |
| $ErFe_2D_{4.2}$ | 5.461(1) | 11.378(2) | 9.364(1) | 122.446(1) | 490.96(9) |

The cell volume variation versus Er content ($\Delta V/\Delta x \approx -12.8(2)$ Å³/Er atom) for both hydride and deuteride series is close to that of their parent intermetallic compounds ($\Delta V/\Delta x = -12.6(2)$ Å³). The average relative cell volume swelling are: $(V_{Hydr.}-V_{Inter.})/V_{Inter.} = 27.5(2)$ % and $(V_{deut.}-V_{Inter.})/V_{Inter.} = 26.7(2)$ % for the hydrides and the deuterides respectively. The cell volume difference between hydride and deuteride for a given Er value is $\Delta V \approx 0.8$ % like that observed for the non-substituted compounds. The larger cell parameters of the hydrides compared to the deuterides confirm the isotope effect already observed in YFe$_2$(H,D)$_{4.2}$ compounds. It has been explained by the larger zero point amplitude of vibration of hydrogen *versus* deuterium atoms inside interstitial sites surrounded by heavy metal atoms [56].

## 2. Magnetic properties

Thermomagnetic curves have been measured for all hydrides and deuterides under an applied magnetic field of 0.03 T. The $M_{0.03T}$ ($T$) curves are compared in Figure 4 for the deuterides with $x = 0$ to 1.

For $x = 0$, only one sharp transition is observed at 84 K corresponding to the FM-AFM transition previously described in [56]. For the Er substituted compounds the magnetization first increases up to a maximum value and then decreases sharply, at a temperature $T_M$, which depends on the Er concentration. ErFe$_2$D$_{4.2}$ shows only a small increase below 10 K, which can be attributed to the onset of the FM-AFM transition. Above $T_M$ a smoother variation is observed, with a slope decrease. According to the previous neutron diffraction studies, the increases of the magnetization from 2 K to $T_M$ is related to the ferrimagnetic behavior, with a decrease of the mean Er moment versus temperature, whereas the mean Fe moment remains almost constant up to $T_M$. The sharp $M$ (T) decrease at $T_M$ is attributed to the ferro-antiferromagnetic transition of the Fe sublattice.

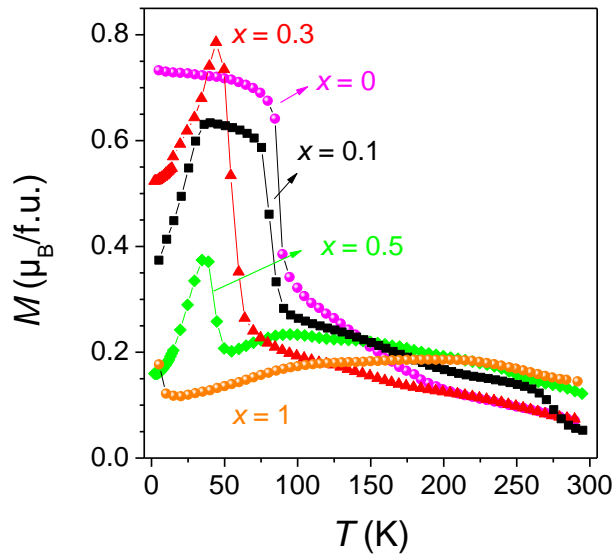

**Figure 4**: Magnetization curves of Y$_{1-x}$Er$_x$Fe$_2$D$_{4.2}$ compounds at $B = 0.03$ T.

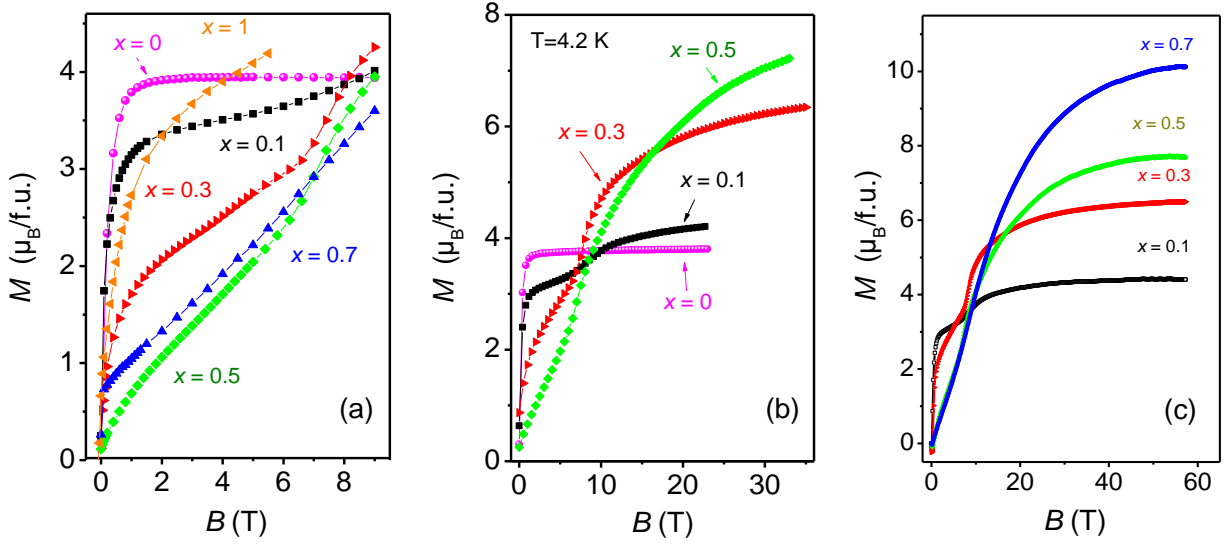

**Figure 5**: $M_T(B)$ curves for $Y_{1-x}Er_xFe_2D_{4.2}$ compounds measured with (a) low magnetic field measurements at 2 K ($x = 0, 0.3, 0.5, 0.7$) and 5 K ($x = 0.1$ and 1), (b) high steady magnetic field at 4.2 K for $x = 0$ to 0.5 and (c) pulsed magnetic field at 4.2 K for $x = 0.1$ to 0.7.

The $M_T(B)$ magnetization curves measured for various deuterides at low temperature ($T \leq 5$ K) on three types of magnetic devices: a) under applied field up to 9 T (PPMS), b) under high static magnetic field up to 23-35 T (LNCMI) and c) under pulsed magnetic fields up to 60 T (HLD) are presented in Figure 5.

For fields below 7 T, the magnetization of Er substituted compounds is smaller than for $YFe_2D_{4.2}$, as expected for a ferrimagnetic behavior. The spontaneous magnetization $M_{spont}$ was obtained for each deuteride by a linear fit of the PPMS data ($2 \leq B \leq 5$ T). $M_{spont}$ first decreases to a minimum at $x \approx 0.5$ and increases again for larger Er content. $M_{spont}$ values were fitted assuming a collinear arrangement of Er, Y and Fe spin sublattices with the equation:

$$M_{spont} = | 2M_{Fe} - x.M_{Er} - (1-x).M_Y | \qquad (2)$$

The best fit was obtained with $M_{Fe} = 2$ $\mu_B$, $M_{Er} = 6.9$ $\mu_B$ and $M_Y = 0.08$ $\mu_B$, and with a chi square $\chi^2$ of 2.65 %. The $M_{Fe} / M_{Er}$ compensation point occurs at $x = 0.57$, indicating an inversion of the Fe orientation parallel and then antiparallel to the applied field. Note that the compensation point is shifted to larger Er concentration than for the parent intermetallic, due to the different values of mean Fe and Er moments.

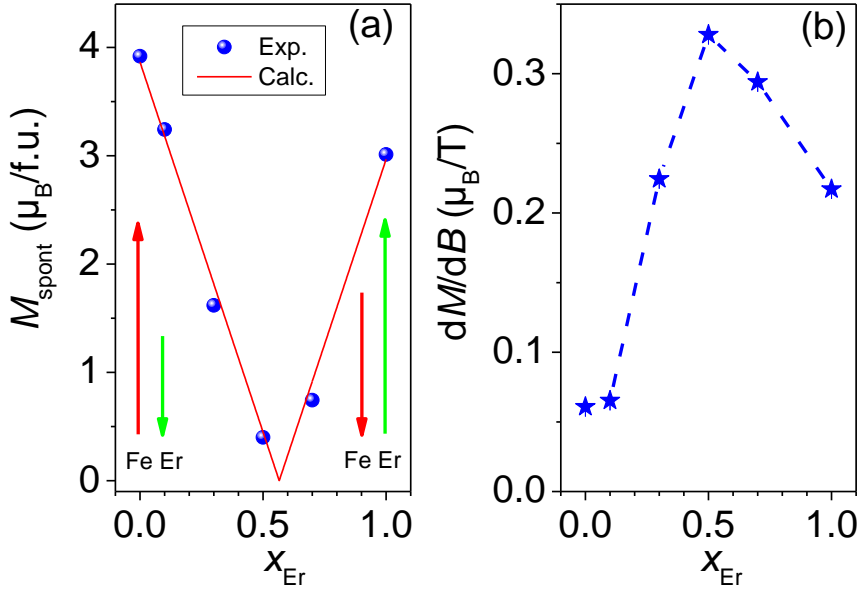

**Figure 6**: (a) Experimental and calculated spontaneous magnetization and (b) d$M$/d$B$ slope for $Y_{1-x}Er_xFe_2D_{4.2}$ compounds at 4.2 K.

Compared to the data obtained for the parent intermetallic compound, one can observe that the Fe moment is larger for the deuteride: 2 $\mu_B$ versus 1.6 $\mu_B$, which can be related to the increase of Fe-Fe distances due to cell volume expansion induced by D insertion. On the other hand, the mean Er moment is reduced from 9 to 6.9 $\mu_B$ whatever the Er concentration. Such reduction of the Er moment compared to the free ion value was already observed by magnetic measurements and neutron diffraction experiments for $Y_{0.7}Er_{0.3}Fe_2$ and $Y_{0.5}Er_{0.5}Fe_2$ compounds [73, 78]. This Er moment reduction has been attributed to a crystal field effect and is observed for all Er substituted compounds. No hysteresis has been observed in the $M_T(B)$ curves contrary to the parent intermetallics, but the d$M$/d$B$ slope changes versus Er concentration (Figure 6b). It displays an opposite variation compared to the $M_{spont}$ (x) curve: it first increases versus rate up to $x = 0.5$ when the total Fe moment is larger than the total Er moment and then slightly decreases. Assuming that, like in $YFe_2D_{4.2}$, the Fe moments are easily aligned parallel to the applied field, the slope is due to a progressive reorientation of the Er moments upon applied field. This effect is maximum at the compensation point. Then when the total Er magnetization becomes larger than the Fe one, the rotation of Er moments requires lower magnetic field.

A deviation from linearity of $M_T(B)$ is observed above 6 T and a field induced transition is observed towards a ferromagnetic alignment of the Fe and Er moments. The transition fields ($B_{trans1}$), estimated from the maximum of the derivative of the $M(B)$ curves, are presented in Figure 7a. The measured values are slightly larger for the pulsed field compared to the steady field approach, and this is probably related to the difference of measurement method, as pulsed field measurement is not strictly at thermal equilibrium. All the transition fields are between 7 and 9 T. They are minimum for $x = 0.5$ and increase again for larger Er contents. It is interesting to observe that the $M(B)$ curves of the Er compounds cross that of YFe$_2$D$_{4.2}$ around 8±0.5 T (Figure 7b).

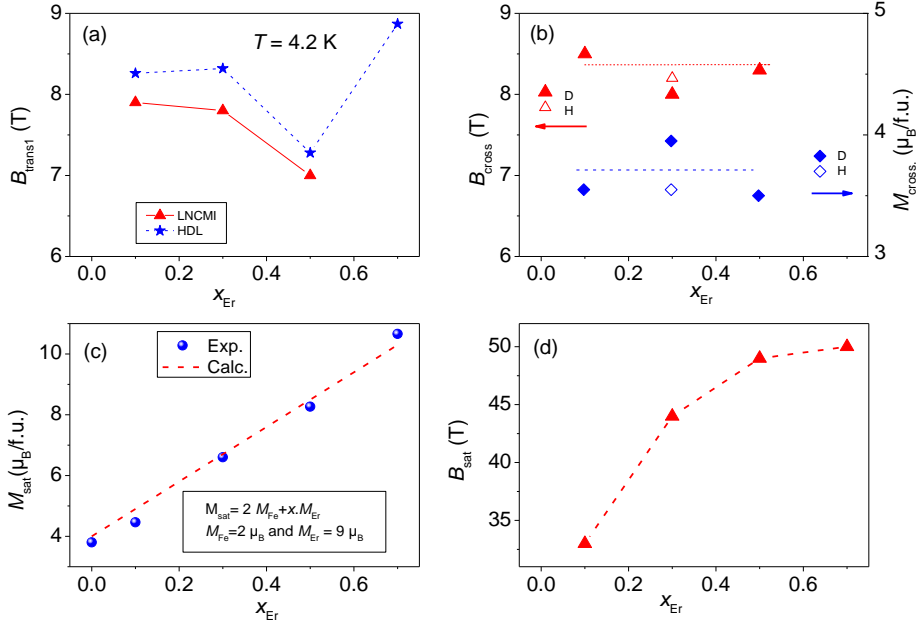

**Figure 7**: (a) Transition fields, (b) crossing fields and magnetizations, c) saturation magnetization (d) saturation fields of Y$_{1-x}$Er$_x$Fe$_2$D$_{4.2}$ compounds.

Above the transition field, the magnetization continues to increase towards a saturation magnetization $M_{sat}$ which has been extrapolated from the pulsed magnetic experiment near 60 T by a $1/B^2$ ferromagnetic approach law (Figure 7c). $M_{sat}$ increases linearly and is well fitted by:

$$M_{sat} = 2 M_{Fe} + x.M_{Er} \qquad (3)$$

with $M_{Fe} = 2$ µ$_B$ and $M_{Er} = 9$ µ$_B$.

This result confirms that the ferromagnetic state can be reached at high field with a mean Er moment corresponding to the free ion value. The field $B_{sat}$ at which the ferromagnetic state is reached, obtained from the pulsed magnetic field experiments, increases from 30 T for $x = 0.1$ to 50 T for $x = 0.7$ (Figure 7d). This evolution can be also attributed to the influence of the Er magnetocrystalline anisotropy, as a larger applied field is necessary to reach the saturation when the Er concentration increases.

The evolution of the magnetization curves $M_T(B)$ at different temperatures is shown for Y$_{0.9}$Er$_{0.1}$Fe$_2$D$_{4.2}$ in Figure 8. Below 45 K, all the $M_T(B)$ curves show an S shape and cross around a magnetic field $B_{cross} = 8.5 \pm 0.5$ T at $M_{cross} = 3.55 \pm 0.5$ µ$_B$/f.u.. The S shape of the $M_T(B)$ curves is progressively smoothed as the temperature increases. At 45 K (Figure 8), the inflection point is no more visible. Similar curve crossing has been observed for the other substituted compounds studied under high magnetic field (Y$_{0.7}$Er$_{0.3}$Fe$_2$D$_{4.2}$, Y$_{0.7}$Er$_{0.3}$Fe$_2$H$_{4.2}$ and $Y_{0.5}Er_{0.5}Fe_2D_{4.2}$) [67, 73, 78] with $B_{cross}$ ranging between 8.0 and 8.6 T, and $M_{cross}$ between 3.5 and 3.95 $\mu_B$/f.u.. Interestingly, the magnetization of $YFe_2D_{4.2}$ is equal to 3.75 $\mu_B$ at 4.2 K for fields between 8 and 8.5 T, corresponding to the midpoint of the crossing magnetizations.

This result supports, as already discussed for the deuterides with $x$ = 0.3 and 0.5 [67, 73, 78], that the Fe sublattice remains ferromagnetic with Fe moments parallel to the applied field as in $YFe_2(H,D)_{4.2}$ compounds. Indeed a canting of the Fe moments was not observed by neutron diffraction. It is rather the Er sublattice which changes upon applied field with a decrease of the Er moment magnetic moment component. In fact, 2 K neutron diffraction experiments have revealed that the Er magnetic Bragg peak intensity goes through a minimum value upon applying field on $Y_{0.5}Er_{0.5}Fe_2D_{4.2}$. A ferrimagnetic-ferrimagnetic transition at $B_{trans1}$ can also be excluded since it is not supported by the present experimental results. Indeed, if the Fe and Er moments were still antiparallel above this field, the magnetization should remain smaller than $2.M_{Fe}$ above $B_{trans1}$ in contradiction with the observed magnetization curves of Figure 5.

This raises therefore the question of the mechanism of the transition from ferrimagnetic towards ferromagnetic state in these pseudobinary compounds. Two main mechanisms can be assumed: a rotation of the Er moment in the ($a$, $c$) basal easy plane with fixed magnetic moments [79] or a demagnetization-remagnetization of the Er moments, which remain collinear to the Fe moments as observed for $Tm_2Fe_{17}$ and $Tm_2Fe_{17}D_{3.2}$ [80, 81]. In this last case, the excited magnetic state evolved upon applying an external magnetic field so that a crossing occurs between the ground state and the excited states leading to zero value of Tm magnetic moment. Note that only macroscopic magnetic properties are reported in the present study, the zero value of the Er magnetic sublattice at the crossing magnetization point gives only the mean value Er magnetization but does not provide information at the microscopic scale. Local spectroscopy at the Er site would be interesting in further studies.

The second mechanism is in better agreement with the experimental results, as in all studied cases the $M_T(B)$ curves are crossing around the same magnetic field and at a magnetization value close to that of the Fe moment in $YFe_2D_{4.2}$ compound. The reduction of the mean Er moment to 6.6 $\mu_B$ in the ferrimagnetic state compared to the Er free ion value of 9 $\mu_B$ as well as the low Er-Fe interaction can favor the proposed mechanism of Er demagnetization.

According to these results, $B_{trans1}$ is not very sensitive to the Er rate and temperature below 50 K, which corresponds to the Er ordering temperature $T_{Er}$ observed by NPD for x = 0.3 ($T_{ER}$ = 55 K) and 0.5 ($T_{ER}$ = 50 K). The comparison of the magnetic properties of $Y_{0.7}Er_{0.3}Fe_2H_{4.2}$ and $Y_{0.7}Er_{0.3}Fe_2D_{4.2}$, has shown that $B_{trans1}$ is also not sensitive to the isotope effect [67]. Several studies on forced ferri-ferromagnetic transitions in $RT_3$, $R_2T_{17}$ intermetallic compounds reported transition fields above 20-40 T [81-85]. Assuming, that the transition field reflects the exchange interactions between $R$ and $T$ moments, its reduction implies a weakening of these interactions.

As hydrogen insertion yields a large increase of the Er-Fe distances in $Y_{0.7}Er_{0.3}Fe_2D_{4.2}$ (around 8 % in average), the indirect exchange interaction between the Er-4f and Fe-3d moments, through the 5d electrons, is strongly reduced. Similar reduction of the $R$-Fe exchange interaction upon hydrogen insertion has been demonstrated by inelastic neutron diffraction experiments on several $R$-$T$ ferrimagnetic compounds like $R_2Fe_{17}$ and $RFe_{11}Ti$ [33, 34] and more recently confirmed on the basis of high magnetic field studies [81, 85, 86]. This can explain why a field of only 8 T is sufficient to reverse the relative orientation of the Er moments in $Y_{1-x}Er_xFe_2D_{4.2}$ compounds as already discussed in refs [73, 78].

The influence of hydrogen insertion on the reduction of transition field has been already investigated in other systems. In $ErFe_{11}TiH$, the H insertion doesn't significantly decrease the field at which a jump of the magnetization occurs - 54 and 56 T for the hydride and its parent compound- but increase the magnitude of the magnetization jump [86]. The jump is also more pronounced along the [001] axis of this tetragonal compound. Calculation based on the crystal field parameters and exchange field of 58.8 T were used to model the magnetization curve up to 200 T at 4.2 K, and reveals that magnetization fields exceeding 60 T are required to reach the full ferromagnetic state. For $Tm_2Fe_{17}D_{3.2}$ only one step like transition is observed around 40 T compared to two transition fields at 48 and 60 T in its parent compound. Thus, the mean molecular field on Tm in $Tm_2Fe_{17}D_{3.2}$ is found to be 49 T or 9% less than in the parent binary compound [81]. The saturated ferromagnetic state can be reached around 60 T for $Tm_2Fe_{17}H_5$ [71]. However the largest influence of hydrogen on such reduction of the transition field for forced ferrri-antiferromagnetic transition is found in $R$Fe$_2$ Hydrides and deuterides ($R$ = Ho, Er, Tm) which present the largest relative cell volume increase. [71]. Our previous studies on $Y_{0.7}Er_{0.3}Fe_2D_{4.2}$ and $Y_{0.5}Er_{0.5}Fe_2D_{4.2}$ have already confirmed this strong reduction of transition field upon deuterium insertion in Laves phase compounds. In this work we can confirm that this effect is present whatever the Er concentration, starting with a very small substituted Er content ($x$ = 0.1)

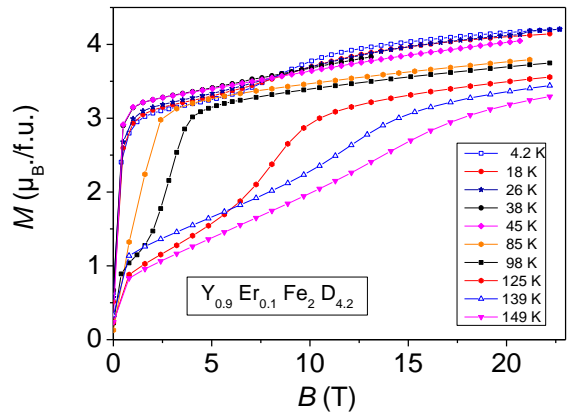

**Figure 8**: Isotherm magnetizations versus field of $Y_{0.9}Er_{0.1}Fe_2D_{4.2}$ measured up to 24 T with high magnetic field device in LNCMI.

As the temperature increases the Er moment is reduced, while that of Fe remains almost constant as observed by NPD for $Y_{0.7}Er_{0.3}Fe_2D_{4.2}$ [73]. At temperatures as high as 45 K, the Er sublattice reaches a paramagnetic state and the metamagnetic transition is no more observed.

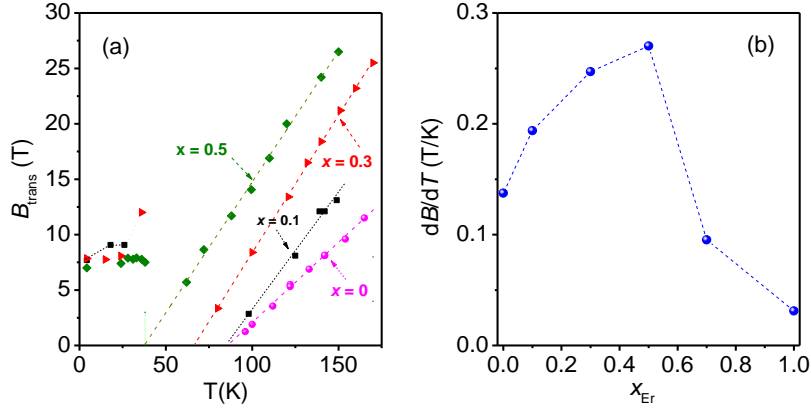

**Figure 9**: (a) transition fields versus temperature for selected $Y_{1-x}Er_xFe_2D_{4.2}$ deuterides and (b) $dB/dT$ slopes for $T > T_{M0}$.

Above $T_M = 50$ K, a second type of metamagnetic behavior is observed for $Y_{0.9}Er_{0.1}Fe_2D_{4.2}$: the $M_T(B)$ curves shows again S-shape characteristic of a metamagnetic behavior, but their inflexion points are shifted to higher fields as the temperature increases. The previous NPD analysis for the deuterides with $x = 0$, 0.3 and 0.5 have shown that it corresponds to a transition from an AFM towards a FM state [73, 77, 78]. The transition fields ($B_{trans2}$), determined from the maximum of the $M_T(B)$ derivative, increases linearly versus temperature. The transition fields $B_{trans2}$ measured for the deuterides are reported versus temperature in Figure 9a and the $dB/dT$ slopes in Figure 9b for deuterides with $x = 0$ to 1. The slope increases from 0 to 0.5 and decrease for larger values. As the compensation point is close to $x = 0.5$, this indicates different $dB/dT$ behaviors depending of the relative contribution of the Er and Fe sublattices.

For all these deuterides as well as the corresponding hydrides, a temperature called $T_{M0}$ can be extrapolated at zero field. In Figure 10a, a linear variation of $T_{M0}$ versus Er concentration is observed and found systematically 43-50 K higher for the hydrides compared to the deuterides. In order to determine whether $T_{M0}$ is related to the cell volume change, $T_{M0}$ has been plotted versus the cell volume measured for both hydrides and deuterides (Figure 10b).

The evolution of $T_{M0}$ versus cell volume displays interesting features: $T_{M0}$ decreases linearly versus the cell volume for both hydrides and deuterides, but a systematic larger temperature of 25 K is observed for the hydrides compared to the deuterides for a given volume. Each parameter (applied pressure, isotope substitution and Er for Y substitution) induces both a cell volume contraction and a linear decrease of $T_{M0}$ versus the cell volume but with different slopes.

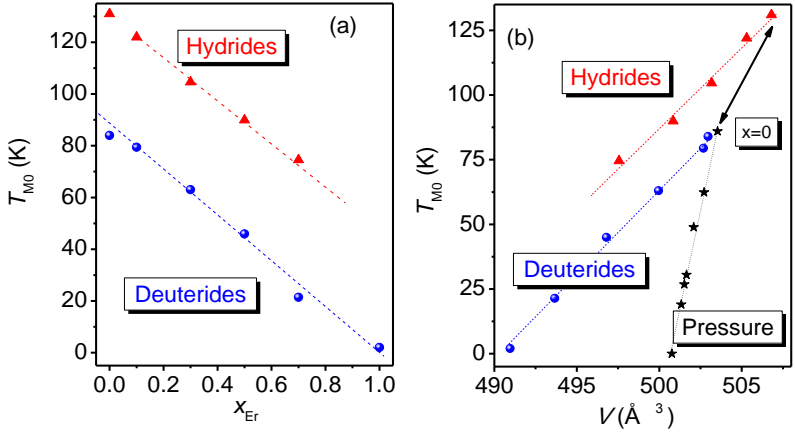

**Figure 10**: Transition temperature $T_{M0}$ versus (a) Er concentration and (b) cell volume for the hydrides and the deuterides of $Y_{1-x}Er_xFe_2$ compounds. The variation of $T_{M0}$ versus cell volume for $YFe_2D_{4.2}$ under pressure (b) is taken from [62].

The slopes $dT_{M0}/dV$= 6.3(2) K/Å$^3$ and 6.7(2) K/Å$^3$ for the Er substituted hydrides and deuterides respectively are about half those measured between $YFe_2H_{4.2}$ and $YFe_2D_{4.2}$ (13.1 K/Å$^3$). In ref [62], the evolution of the structural and magnetic properties of $YFe_2D_{4.2}$ has been investigated under pressure. The variation of the cell volume versus pressure $dV/dP$ =-4.76 Å$^3$ GPa$^{-1}$ combined to the variation of $T_{M0}$ versus pressure $dT_{M0}/dP$ = -156 K/GPa leads to $dT_{M0}/dV$= 16.4 K/Å$^3$.

These results are very interesting as they clearly show that the metamagnetic transition is not simply driven by the cell volume variation and that the influence of the Er substitution as well as the H for D replacement on $T_{M0}$ cannot be considered as a simple chemical pressure effect.

Considering that the applied pressure is a pure volume effect, we should notice that $dT_{M0}/dV$ is different for the hydride and the deuteride but converge to a critical volume $V_0 \approx 500$ Å$^3$. It means that below a critical mean Fe-Fe distance, the AFM structure becomes more stable than the FM one. But when Y is substituted by Er, the FM state is still more stable for lower cell volumes, meaning that the Er contribution can stabilize a ferromagnetic state for lower Fe-Fe distances. The Er substitution should modify the DOS at $E_F$ and can also contribute by acting as a magnetic molecular field.

A different influence of chemical substitution and applied pressure effect between two ferromagnetic phases (FM1 and FM2) and FM2 to paramagnetic transition temperature has been observed on the magnetic transition of $Sc_{1-x}Ti_xFe_2$ [87]. This difference was explained by an alloying effect which modifies the electronic state rather than a volume effect. In the $Y_{1-x}R_xFe_2$ hydrides and deuterides there is probably a competition between the volume contraction induced by the smaller size of the Er ion compared to Y ion and electronic effects, related to the contribution of the 5d electrons of the rare earth to the DOS. Several works on itinerant electron metamagnetism have suggested the influence of the shape of the density of state at the Fermi level. In addition, such transition can be also influenced by the existence of large spin fluctuations on the transition metal sublattice. Each Fe atom is surrounded by 4 to 5 H or D neighbors which modify the shape of the density of state [88]. As the H and D atoms present different zero-point amplitudes of vibration and additionally a dependence of the phonon contribution versus temperature, they can also strongly influence the Fe sublattice metamagnetic behavior.

## IV. Conclusions

It has been evidenced, in this study, how hydrogen and deuterium insertion can strongly modify both structural and magnetic properties of $Y_{1-x}Er_xFe_2$ compounds. For H(D) concentration of 4.2 H(D)/f.u. a lowering of the crystal structure symmetry from cubic to monoclinic structure and a cell volume increase around 26-27 % are observed. At low temperatures (below 50 K) a forced ferri-ferromagnetic transition is found at a transition field of only 8 T, due to the reduction of the Er-Fe interactions by H insertion. The transition field is not sensitive to the Er composition nor the isotope effect, but as the Er concentration increases a larger applied field is necessary to reach ferromagnetic saturation due to the magnetocrystalline anisotropy of the Er moment.

Another metamagnetic process is observed being related to an itinerant electron metamagnetic behavior of the Fe sublattice and corresponding to a FM-AFM transition. This transition is highly sensitive to any structural change induced by Er for Y substitution, D for H replacement and external applied pressure. Although all these effects lead to a reduction of the cell volume, they have not the same influence on the onset temperature $T_{M0}$, above which the metamagnetic transition occurs. This implies that the FM-AFM transition is not simply tuned by cell volume changes, but also by the modifications of the density of electron near the Fermi level upon Y for Er chemical substitution. A significant (H,D) isotope effect has been also found, which can be related to the influence of phonons on the transition temperature.

Further experimental studies on the magnetic properties of these pseudobinary compounds under external applied pressures are in progress to discriminate the influence of electronic and volume effects on both transitions.

**Acknowledgments**

We are thankful to T. Leblond for the synthesis of the intermetallic compounds and some of the hydrides and deuterides during its PhD Thesis. We thank the high magnetic field laboratories of Grenoble (LNCMI-CNRS) and Dresden (HLD-HZDR), members of the European Magnetic Field Laboratory (EMFL) for the time given to perform the magnetic measurements under static and pulsed magnetic field respectively.